\documentclass[aps,prb,twocolumn,superscriptaddress,showpacs]{revtex4}
\usepackage{graphicx}
\usepackage{latexsym}
\usepackage{amssymb}
\usepackage{amsmath}
\usepackage{amsfonts}
\usepackage{bm}
\usepackage{multirow}
\usepackage{color}
\usepackage{comment}
\newcommand{\ii}{\mathrm{i}}

\newcommand{\SO}{\mathrm{SO}}

\newcommand{\SU}{\mathrm{SU}}
\newcommand{\U}{\mathrm{U}}

\newcommand{\vect}[1]{{\bm{#1}}}

\newcommand{\beq}{\begin{equation}}
\newcommand{\eeq}{\end{equation}}
\newcommand{\beqn}{\begin{eqnarray}}
\newcommand{\eeqn}{\end{eqnarray}}

\DeclareMathAlphabet{\mathbbold}{U}{bbold}{m}{n}

\def\SU{{\rm SU}}

\def\U{{\rm U}}

\def\ra{\rightarrow}

\begin{document}

\title{Generic ``Unnecessary" Quantum Critical Points with Minimal Degrees of Freedom}

\author{Chao-Ming Jian}
\affiliation{Kavli Institute of Theoretical Physics, Santa
Barbara, CA 93106, USA}

\author{Cenke Xu}
\affiliation{Department of Physics, University of California,
Santa Barbara, CA 93106, USA}

\begin{abstract}

We explore generic ``unnecessary" quantum critical points with
minimal degrees of freedom. These quantum critical points can be
avoided with strong enough symmetry-allowed deformations of the
Hamiltonian, but these deformations are irrelevant perturbations
below certain threshold at the quantum critical point. These
quantum critical points are hence {\it unnecessary}, but also {\it
unfine-tuned} (generic). The previously known examples of such
generic unnecessary quantum critical points involve at least eight
Dirac fermions in both two and three spatial dimensions. In this
work we seek for examples of generic unnecessary quantum critical
points with minimal degrees of freedom. In particular, in three
dimensional space, we identify two examples of such generic
unnecessary quantum critical points. The first example occurs in a
$3d$ interacting topological insulator, and it is described by
{\it two} $(3+1)d$ massless Dirac fermions in the infrared limit;
the second example occurs in a $3d$ topological superconductor,
and it is formally described only {\it one} $(3+1)d$ massless
Dirac fermion.

\end{abstract}

\maketitle

\section{Introduction}

A quantum critical point (QCP) is usually found between two
different phases of matter with qualitatively different
properties. A generic QCP has only one symmetry allowed relevant
perturbation, which is the tuning parameter of the quantum phase
transition, in other words a generic QCP is ``unfine-tuned". In
fact, the existence of a generic QCP in a phase diagram usually
implies that the two phases on the opposite sides of the QCP are
both stable fixed points under renormalization group (RG), and the
QCP is unavoidable (or ``necessary"), meaning there does not exist
a smooth adiabatic route in the phase diagram that connects the
two sides of the QCP, no matter how the Hamiltonian is deformed as
long as certain symmetry is preserved. The simplest example that
illustrates this common wisdom is the transition of the quantum
Ising model in any spatial dimension. This Ising QCP is sandwiched
between a disordered phase which preserves all the symmetries, and
an ordered phase with spontaneous Ising ($Z_2$) symmetry breaking.
The Ising QCP is ``necessary", meaning if one tries avoiding this
QCP by deforming the Hamiltonian, the best one can do is to drive
the QCP into a first order transition across a tricritical point,
$i.e.$ there is no smooth route that connects the disordered and
ordered phases.

\begin{figure}
\includegraphics[width=0.9\linewidth]{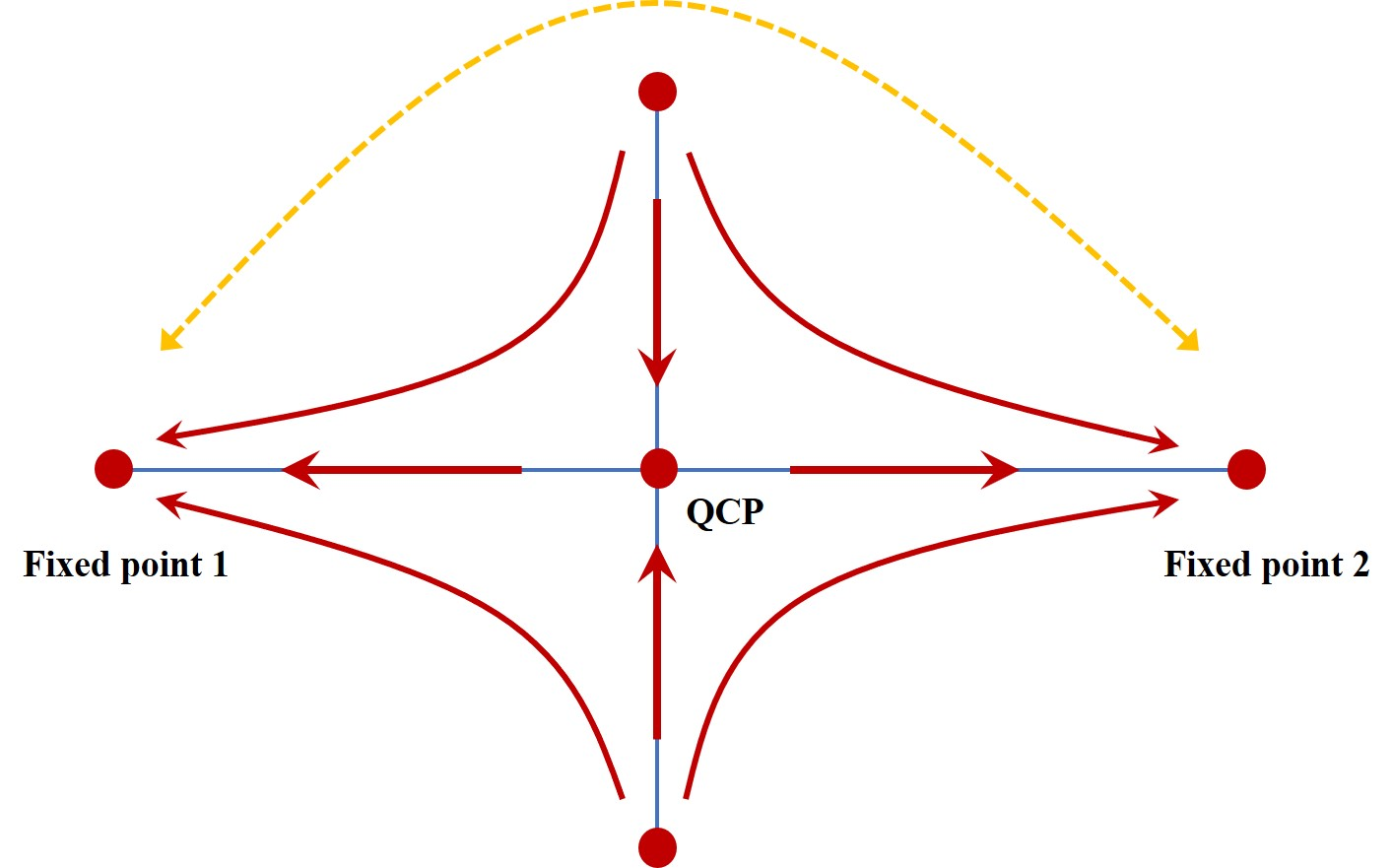}
\caption{ The schematic phase diagram and renormalization group
flow of the generic unnecessary QCP. The horizontal axis is the
tuning parameter between the two phases (stable fixed points 1 and
2), in the examples discussed in this work it corresponds to the
mass of either two or one $(3+1)d$ Dirac fermions; the vertical
axis is the interaction, which is perturbatively irrelevant at the
QCP in the noninteracting limit. There exists a adiabatic curve
(dashed curve) in the entire phase diagram that connects fixed
points 1 and 2. } \label{AQCP}
\end{figure}

The Ising QCP is a quantum analogue of the classical Ising
transition, and it is sandwiched between two phases with classical
analogues at finite temperature. The study of quantum many-body
systems have revealed that the quantum phases are far richer than
classical phases, examples include the topological
phases~\cite{wenreviewtopo} and symmetry protected topological
phases~\cite{wenspt,wenspt2} (such as topological insulators),
which in the Landau-Ginzburg paradigm all correspond to the same
disordered phase. Then we may need to revisit the wisdom we
learned from classical critical phenomena as well. Exotic QCPs
beyond the classic Landau-Ginzburg-Wilson-Fisher paradigm have
been extensively discussed both
theoretically~\cite{deconfine1,deconfine2} and
numerically~\cite{deconfinesandvik1,deconfinesandvik2,deconfinemelko},
including recently developed duality understanding of these
QCPs~\cite{deconfinedual,deconfinedualnumeric}. Mostly recently,
new possibilities of QCPs have been pointed out~\cite{bisenthil}:
there are {\it generic unfine-tuned} but meanwhile {\it
unnecessary} (avoidable) QCPs, namely there exists a symmetry
allowed route which corresponds to a strong enough deformation of
the Hamiltonian in the phase diagram that smoothly connects the
two stable fixed points (phases) on the two sides of the QCP, but
the deformation is perturbatively irrelevant at the QCP below
certain threshold. The schematic phase diagram and RG flow around
the unnecessary QCP is sketched in Fig.~\ref{AQCP}.

The simplest generic unnecessary QCP examples discussed in
Ref.~\onlinecite{bisenthil} require eight massless Dirac fermions
with a $\SO(7)$ symmetry, in both two and three spatial
dimensions. This phenomenon is deeply related to the interacting
topological insulators, and it was understood that interaction can
reduce/collapse the classification of some of the topological
insulator (TI) or topological superconductor
(TSC)~\cite{fidkowski1,fidkowski2,chenhe3B,senthilhe3,youinversion,qiz8,zhangz8,yaoz8,levinguz8}.
For example, the TSC $^3$He-B phase has a $\mathbb{Z}$
classification with time-reversal symmetry with $\mathcal{T}^2 =
-1$, but under time-reversal allowed interaction the
classification of this TSC is reduced to $\mathbb{Z}_{16}$. This
implies that for 16 copies of the TSC, the topological nontrivial
and trivial phases in the noninteracting limit can be connected
smoothly with strong enough local interaction. However, in the
noninteracting limit, the topological-to-trivial transition of 16
copies of TSC is described by 16 massless Majorana fermions (or
mathematically equivalent to 8 massless Dirac fermions), then as
long as we impose an extra flavor symmetry (for example the
$\SO(7)$ symmetry in Ref.~\onlinecite{bisenthil}) to guarantee
that the 16 Majorana fermions all become massless simultaneously
while the reduction of classification is still valid with the
extra flavor symmetry, then this QCP in the noninteracting limit
is still a generic unfine-tuned QCP, because short range
interactions are irrelevant at the free massless Dirac fermion
fixed point for spatial dimensions higher than $1$.

In this work we seek for much simpler examples of such generic
unnecessary QCPs. We discuss two examples which happen in strongly
interacting $3d$ TI and $3d$ TSC respectively. In the infrared
limit, the first example is described by only {\it two} massless
$(3+1)d$ Dirac fermion. In the noninteracting limit, one side of
the QCP is a trivial state, while the other side is a $3d$
topological insulator with $\U(1)\times Z_n \times \mathcal{P}$
symmetry, where $n$ is an odd integer, and $\mathcal{P}$ is a
spatial-reflection. In the noninteracting limit, this TI has a
$\mathbb{Z}$ classification, and the trivial-to-topological
transition is described by two $(3+1)d$ massless Dirac fermions.
An extra time-reversal symmetry $\mathcal{T}$ with $\mathcal{T}^2
= +1$ will guarantee that there is only one direct transition
between the trivial and topological insulator phases. However, we
will demonstrate that this TI can always be trivialized by local
interactions, hence the QCP of the trivial-to-topological
transition in the noninteracting limit can be avoided by strong
enough interaction, which is perturbatively irrelevant at the QCP
in the nonnteracting case.

The second example of unnecessary QCP we find is even simpler: it
is formally described by only {\it one} massless Dirac fermion
(two massless Majorana fermions) in the infrared limit, and one
side of the QCP is a $3d$ TSC with $(Z_{2n} \rtimes Z_4^T)/Z_2$
symmetry with an odd integer $n$ (the $Z_4^T$ refers to a
time-reversal symmetry with $\mathcal{T}^2 = -1$), while the other
side is a trivial superconductor. In the noninteracting limit, the
TSC is topologically nontrivial with a classification
$\mathbb{Z}_2$, while this TSC is again trivialized by local
interactions. Hence the trivial-to-topological transition is again
unnecessary but generic.

\section{Generic Unnecessary QCP in a $3d$ TI}

\subsection{Preparation: $2d$ TI with $\U(1)\times Z_n$ symmetry}

To construct our $3d$ system, we need to first understand the $2d$
TI with the $\U(1)\times Z_n$ symmetry, and we will focus on the
case with odd integer $n$. This TI was discussed in
Ref.~\onlinecite{ashvinzn,vci} in different physics contexts, and
for $n = 3$ this TI corresponds to the valley Chern
insulator~\cite{vci} that is realized in Moir\'{e}
systems~\cite{band3,band5,TDBGt,dai2,hall,FM,zaletel,zhangmao}. In
this section we will review the understanding of interacting $2d$
TI with $\U(1)\times Z_n$ symmetry. We will also impose another
time-reversal symmetry $\mathcal{T}$ with $\mathcal{T}^2 = +1$,
hence the entire symmetry group is $(\U(1) \rtimes Z_2^T) \times
Z_n$. This time-reversal symmetry $\mathcal{T}$ is realized in the
spin-polarized correlated insulator at half-filling in the
miniband of the twisted double biayer graphene
system~\cite{kimtalk,TDBG1,TDBG2,TDBG3}, and $\mathcal{T}$
corresponds to an ordinary time-reversal symmetry of electrons
times a spin flipping.

This TI can be naturally embedded into a nonchiral topological
insulator (TI) with $\U(1)_c \times \U(1)_s$ symmetry. There are
only two elementary fermions with charge $(1,1)$ and $(1,-1)$
under the $\U(1)_c\times \U(1)_s$ symmetry, and for the simplest
case they form Chern insulators with Chern number $\pm 1$
respectively. At the free fermion level, the Hamiltonian of the
$1d$ edge state of this TI is \beqn H = \int dx \ \psi^\dagger
\left( - \ii \sigma^3
\partial_x \right) \psi. \label{edge} \eeqn The symmetries act on the
boundary fermions as \beqn \U(1) &:& \psi_1 \ra e^{\ii
\alpha}\psi_1, \ \ \ \psi_2 \ra e^{\ii \alpha}\psi_2 \cr\cr Z_n
&:& \psi_1 \rightarrow e^{2\pi \ii /n}\psi_1,  \ \ \ \psi_2
\rightarrow e^{-2\pi \ii /n}\psi_2. \cr\cr \mathcal{T} &:& \psi_1
\ra \psi_2, \ \ \ \psi_2 \ra \psi_1. \eeqn The $Z_n$ symmetry
guarantees that no fermion bilinear mass term can be added to the
boundary Hamiltonian. Also, for arbitrary copies of the TI,
fermion bilinear mass operators are always forbidden. Hence the
classification of this TI in the noninteracting limit is
$\mathbb{Z}$.

To describe the nonchiral TI {\it with ininteraction}, we can use
the $K-$matrix formalism~\cite{wenzee}. The system can be
described by the following Chern-Simons theory \beqn \mathcal{L} =
\frac{\ii}{4\pi} \sum_{A,B=1,2} K^{AB} a^A \wedge d a^B, \ \ \ K =
\left(
\begin{array}{cc}
1 & 0 \\
0 & -1
\end{array}
\right), \eeqn where $a^A$ with $A=1,2$ are two dynamical U(1)
gauge fields. The edge state of this TI is described by the
Luttinger liquid theory with two chiral boson fields $\phi_{1}$,
$\phi_2$ and the same $K-$matrix above~\cite{wenedge,wenreview}:
\begin{align}
\mathcal{L}_{\rm edge} = \sum_{A,B=1,2}\frac{K^{AB}}{4\pi}
\partial_x \phi_A \partial_t \phi_B - \frac{V^{AB}}{4\pi}
\partial_x \phi_A \partial_x \phi_B,
\label{Eq:Edge_Luttinger_Luquid}
\end{align}
where $V$ is a $2\times 2$ positive-definite velocity matrix. In
this theory, the boson fields satisfy the equal time commutation
relation $[\phi_A(x), \partial_y \phi_B(y)]= 2\pi \ii
(K^{-1})^{AB} \delta(x-y)$. Under the $\U(1) \times Z_n$ and
time-reversal symmetry, the chiral boson fields $\phi_{1,2}$
transform as \beqn \U(1) &:&  \phi_{1} \rightarrow \phi_{1} +
\alpha, \ \ \phi_{2} \rightarrow \phi_{2} + \alpha, \cr\cr Z_n &:&
\phi_{1} \rightarrow \phi_{1} + \frac{2\pi}{n}, \ \ \phi_2 \ra
\phi_2 - \frac{2\pi}{n}, \cr\cr \mathcal{T} &:& \phi_1 \ra -
\phi_2, \ \ \phi_2 \ra - \phi_1. \eeqn

Now we demonstrate that the nonchiral TI with $\U(1) \times Z_n$
and the time-reversal symmetry defined above at most has a
$\mathbb{Z}_n$ classification under local interaction, for odd
integer $n$. The fact that $n$ copies of such TI together are
topologically trivial can be seen from the edge theory of this
system which consists of $n$ copies of Luttinger liquid theory Eq.
\ref{Eq:Edge_Luttinger_Luquid}. Let's denote chiral boson fields
in this $n$-copy Luttinger liquid theory as $\phi_{i,A}$ where
$i=1$ is the copy index and $A=1,2$ is the label for the chiral
bosons within each copy. The boundary of $n$ copies of the TI can
be gapped out by the following symmetric boundary interaction
without ground state degeneracy: \beqn \mathcal{L}_{\rm
edge}^{(1)} &=& - \cos\left( \sum_{i = 1}^n (\phi_{i,1} -
\phi_{i,2}) \right) \cr\cr &-& \sum_{i = 1}^{n-1} \cos(\phi_{i,1}
+ \phi_{i,2} - \phi_{i+1,1} - \phi_{i+1,2}). \label{Ledge1}\eeqn
These are local interacting terms between the electrons, and they
preserve all the symmetries.

This edge theory $\mathcal{L}_{\rm edge}^{(1)}$ can be analyzed
systematically as following: There are in total $n$ different
terms in $\mathcal{L}_{\rm edge}^{(1)}$, and we can represent each
term in $\mathcal{L}_{\rm edge}^{(1)}$ as $\cos(\mathbf{\Lambda}_I
\cdot \mathbf{\Phi})$. $\mathbf{\Lambda}_I $ are $2n$ component
vectors ($I = 1, \cdots n$), and $\mathbf{\Phi} = (\phi_{1,1},
\phi_{1,2}, \phi_{2,1} \cdots)$. $\mathbf{\Lambda}_I$ are a set of
minimal linearly independent integer vectors, and they satisfy the
condition~\cite{nullhaldane,nulllevin} \beqn \mathbf{\Lambda}^t_I
\mathbf{K}^{-1} \mathbf{\Lambda}_J = 0, \label{null} \eeqn for any
$I, J = 1, \cdots n$. Here $\mathbf{K}$ is the $2n \times 2n$
block-diagonal $K-$matrix for $n$ copies of the nonchiral TI with
$\U(1) \times Z_n$ symmetry. Eq.~\ref{null} implies that the
arguments in all the cosine terms in $\mathcal{L}_{\rm
edge}^{(1)}$ commute with each other, and hence all terms in
$\mathcal{L}_{\rm edge}^{(1)}$ can be minimized simultaneously,
which leads to a fully gapped edge state.

The $2n-$component integer vector $\mathbf{\Lambda}_I$ is a vector
in a $2n$ dimensional cubic lattice with lattice constant $1$.
Linear combinations of $\mathbf{\Lambda}_I$ span a $n-$dimensional
hyperplane of this $2n$ dimensional cubic lattice. To be rigorous
we also need to show that $\mathbf{\Lambda}_I$ are the irreducible
basis vectors of the lattice sites residing on this
$n-$dimensional hyperplane, hence the minimum of $\mathcal{L}_{\rm
edge}^{(1)}$ has no degeneracy. This can also be verified for odd
integer $n$.

Having shown that $n$ copies of the nonchiral TI with $(\U(1)
\rtimes Z_2^T) \times Z_n $ symmetry together is topologically
trivial, we now argue that the classification of such nonchiral TI
has to be $\mathbb{Z}_n$, with odd integer $n$.  For $k$ copies of
such nonchiral TI, we can consider, in the non-interacting limit,
$k$ copies of the edge theory Eq. \ref{edge} residing on a circle.
Every time a $2\pi$ flux associate to the U(1) symmetry is
threaded through the circle, the total $Z_n$ charge of this
$k$-copy edge theory is shifted by $2k$.  Given that $Z_n$ charges
are defined modulo odd integer $n$, if the number of copies $k$ is
not a multiple of $n$, the shift of $Z_n$ charge on the edge is
non-trivial indicating that the edge theory is in fact anomalous
and further suggesting that its associated bulk is topologically
non-trivial. Therefore, $n$ is the ``minimal number" of copies
needed for the edge to be non-anomalous. Combined with the
previous argument, the classification of the nonchiral TI with
$(\U(1) \rtimes Z_2^T) \times Z_n $ symmetry has to be
$\mathbb{Z}_n$.

For odd integer $n$, a general connection between the nonchiral TI
with $(\U(1) \rtimes Z_2^T) \times Z_n $ symmetry and bosonic
symmetry protected topological (bSPT) state with the same
symmetry~\cite{wenspt,wenspt2} can be made. Since the interacting
TI has a $\mathbb{Z}_n$ classification, one copy of the elementary
TI is topologically equivalent to $n+1$ (an even integer) copies
of the TI; while according to
Ref.~\onlinecite{xugraphene,spn,xufb,hermelecrystal}, even number
of such nonchiral TIs can be ``glued" into a bSPT state with the
same symmetry under interaction, where all the local fermion
excitations at the boundary are gapped out by interaction, leaving
only symmetry protected gapless local bosonic excitations.

A variety of bSPT states and their edge states can be described by
the Chern-Simons theory with the following
$K$-matrix~\cite{luashvin}, whose boundary state is described by
two chiral bosons $\varphi$ and $\theta$ with $K-$matrix:
\begin{align}
K_{\rm bSPT} = \left(
\begin{array}{cc}
0 & 1 \\
1 & 0
\end{array}
 \right).
 \label{Eq:K_bSPT}
\end{align}
The chiral bosons transform under the symmetries as \beqn \U(1)
&:& \varphi \rightarrow \varphi + 2\alpha, \ \ \ \theta
\rightarrow \theta \nonumber \cr\cr Z_n &:& \varphi \rightarrow
\varphi, \ \ \ \theta \rightarrow \theta - 2\pi /n, \cr\cr
\mathcal{T} &:& \varphi \ra - \varphi, \ \ \ \theta \ra \theta.
\label{Eq: symmetry_action_bSPT_edge} \eeqn Now, we consider a
$(1+1)d$ interface between the bSPT and the nonchiral TI discussed
previously which can be described by the four boson fields
$\phi_{1,2}$, $\varphi$ and $\theta$. The symmetry allowed
interaction that can gap out this interface without degeneracy is
\beqn \mathcal{L}_{\rm edge}^{(2)} \sim - \cos(\phi_1 + \phi_2 -
\varphi) - \cos(\phi_1 - \phi_2 + 2\theta).
\label{Eq:QSH-BSPT_GappedInterface} \eeqn Again the arguments in
the cosine terms commute with each other, hence all terms in
$\mathcal{L}_{\rm edge}^{(2)}$ can be minimized simultaneously,
and the interface is gapped out without degeneracy through the
same reasoning as before. The existence of such a gapped interface
between the bSPT and the fermionic TI guarantees the topological
equivalence between the two sides of this interface.

The physical interpretation of the bosonoic fields $\varphi$ and
$\theta$ can be understood in terms of their quantum numbers. The
local boson field $e^{\ii \varphi}$ can be identified with the
bound state $\psi_1 \psi_2$;
the quantum number of the single boson operator $e^{\ii \theta}$
is equivalent to $(n+1)/2$ copies of the particle-hole pair
$\psi_1^\dag\psi_2$ ($Z_n$ charge is defined mod $n$). We can see
that {\it when and only when} $n$ is an odd integer, $e^{\ii
\theta}$ can be viewed as a local boson field. Hence for odd
integer $n$, a nonchiral TI with $\U(1) \times Z_n $ symmetry and
time-reversal $\mathcal{T}$ with $\mathcal{T}^2 = +1$ is
equivalent to a bSPT constructed with local bosons.

This result implies that, under interaction the $2d$ nonchiral TI
with $(\U(1)\rtimes Z_2^T) \times Z_n$ symmetry can have fully
gapped single electron excitation, but meanwhile symmetry
protected gapless local boson excitations at its boundary. This
was thought to be only possible for {\it even} copies of nonchiral
TI such as the quantum spin Hall insulator with spin $S^z$
conservation~\cite{xugraphene,spn,xufb,hermelecrystal}.

\subsection{Generic unnecessary QCP in two dimensions}

The result in the previous section is sufficient to predict a
generic unnecessary QCP in $2d$. In the noninteracting limit, the
topological transition between the trivial state and the nonchiral
TI with $(\U(1) \rtimes Z_2^T) \times Z_n$ symmetry is described
by two massless $(2+1)d$ two-component Dirac fermions. The
time-reversal symmetry $\mathcal{T}$ guarantees that there is a
single transition between the trivial and topological state. Now
if we consider $n$ copies of the TIs, with extra assumptions of a
discrete cyclic symmetry between the $n$ copies of TIs, and also a
fermion parity of each copy of the TI, a single generic direct
trivial-to-topological transition of $n$ copies of the TIs is
still guaranteed, which is described by $2n$ massless $2d$ Dirac
fermion.

One can check that the interaction Lagrangian Eq.~\ref{Ledge1}
also preserves the extra cyclic symmetry and fermion parity of
each copy, then $n$ copies of the TI is still trivialized by
interaction. Though short range interaction is irrelevant at the
$(2+1)d$ Dirac fermion, the generic direct trivial-to-topological
transition of $n-$copies of the TI described above becomes
unnecessary under strong enough symmetry-allowed local
interaction. The minimum situation we find in $2d$ would be $n =
3$, which may have been realized as the spin polarized correlated
insulator in the twisted double bilayer
graphene~\cite{kimtalk,TDBG1,TDBG2,TDBG3}.

In this section we demonstrate that the generic unnecessary QCP
can be even simpler in $3d$. In the noninteracting limit, the
topological transition between the trivial state and the
nontrivial $3d$ TI with $\U(1) \times Z_n \times \mathcal{P}$
symmetry is described by two massless $(3+1)d$ four-component
Dirac fermions. But we will demonstrate that this transition is
unnecessary under interaction.

\subsection{Noninteracting $3d$ TI with $\U(1)\times Z_n \times \mathcal{P}$ symmetry}

Now we switch gear to the $3d$ TI with $\U(1)\times Z_n \times
\mathcal{P}$ symmetry. The $2d$ boundary state of this TI is
described by the Hamiltonian \beqn H_{\mathrm{edge}} = \int d^2x \
\psi^\dagger ( \ii \sigma^{10}
\partial_x + \ii \sigma^{33} \partial_y )\psi, \label{2dedge} \eeqn
$\sigma^{ab} = \sigma^a \otimes \sigma^b$. The symmetries act on
the $2d$ boundary fermions as \beqn \U(1) &:& \psi \ra e^{i\alpha}
\psi, \cr\cr \mathcal{P} &:& x \rightarrow -x, \ \ \psi
\rightarrow \sigma^{30} \psi, \cr\cr Z_n &:& \psi \rightarrow \exp
\left(\ii \frac{2\pi}{n} \sigma^{03} \right) \psi. \eeqn One can
check that all the mass terms at this $2d$ boundary, such as \beqn
\psi^\dagger \sigma^{20} \psi, \ \ \psi^\dagger \sigma^{23} \psi,
\ \ \psi^\dagger \sigma^{31} \psi, \ \ \psi^\dagger \sigma^{32}
\psi, \eeqn are forbidden by either the reflection, or the $Z_n$
symmetry. We can also add another time-reversal symmetry
$\mathcal{T}$ with $\mathcal{T}^2 = +1$: \beqn \mathcal{T} : \psi
\rightarrow \sigma^{31} \psi. \eeqn Also, one can check that for
arbitrary copies of the system, all the fermion bilinear mass
terms are still forbidden by either the $\mathcal{P}$ or $Z_n$
symmetry, hence the classification of this $3d$ TI in the free
fermion limit is $\mathbb{Z}$.

At the free fermion level, the bulk trivial-to-topological phase
transition of a single copy of this $3d$ TI is described by the
following Hamiltonian: \beqn H_{\mathrm{bulk}} &=& \int d^3x \
\psi^\dagger \left( \ii \sigma^{103}
\partial_x + \ii \sigma^{333}
\partial_y + \ii \sigma^{002}\partial_z \right)\psi \cr\cr &+& m \psi^\dagger
\sigma^{001}\psi. \label{3dbulk} \eeqn Now $\psi$ becomes an eight
component fermion. All the symmetries act on the bulk fermions as
\beqn \mathcal{P} &:& \psi \rightarrow \sigma^{300} \psi, \ \ \
\mathcal{T}: \psi \rightarrow \sigma^{310} \psi, \cr\cr \U(1) &:&
\psi \ra e^{i \alpha} \psi, \ \ Z_n : \psi \rightarrow \exp
\left(\ii \frac{2\pi}{n} \sigma^{030} \right) \psi. \eeqn One can
check that, at the $xy$ $2d$ boundary of the system, the bulk
Hamiltonian Eq.~\ref{3dbulk} reduces to the $2d$ boundary
Hamiltonian Eq.~\ref{2dedge}, which corresponds to a domain wall
of $m$ long the $\hat{z}$ axis in Eq.~\ref{3dbulk}, and all the
bulk symmetry actions reduce exactly to the symmetry actions on
the boundary defined above.

There are other mass matrices in the $3d$ bulk: \beqn
\sigma^{203}, \ \ \sigma^{233}, \ \ \sigma^{313}, \ \
\sigma^{323}, \ \ \sigma^{121}, \ \ \sigma^{111}, \ \
\sigma^{031}. \eeqn All these extra mass terms are forbidden, one
way or another. The most interesting, and probably important
``extra" mass term is the last one: $\psi^\dagger \sigma^{031}
\psi$. This mass term, if exists, does not gap out the transition,
but split the transition into two. But this extra mass term is
forbidden by $\mathcal{T}$. So $\mathcal{T}$ is the symmetry that
guarantees a generic direct single trivial-to-topological
transition at the free fermion level. But later we will show that
this transition will be avoided under interaction.

\subsection{Interacting TI and Unnecessary QCP in three dimensions}

\begin{figure}
\includegraphics[width=0.8\linewidth]{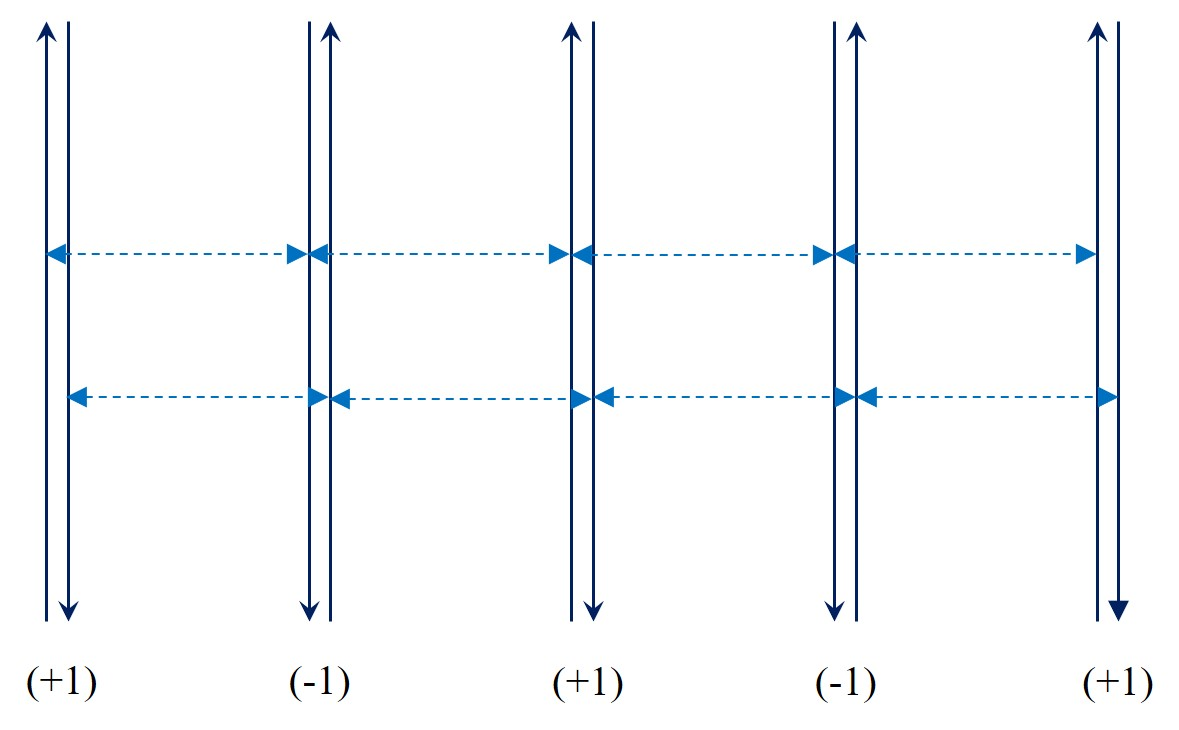}
\caption{The coupled wire construction of the $2d$ edge state
Hamiltonian Eq.~\ref{2dedge} of the noninteracting $3d$ TI with
$\U(1) \times Z_n \times \mathcal{P} $ symmetry. Each wire
represents the $1d$ edge state of a $2d$ TI with $\U(1) \times
Z_n$ symmetry, which is described by Eq.~\ref{edge}. Each wire is
composed of a pair of counter-propagating $1d$ fermion modes that
carry opposite $Z_n$ charges. Symmetry allowed tunnellings between
the wires will drive the system into the $2d$ edge state
Hamiltonian Eq.~\ref{2dedge}. } \label{wire1}
\end{figure}

In Ref.~\onlinecite{hermelep}, a general approach of understanding
and constructing $3d$ symmetry protected topological (SPT) state
(generalization of topological insulator) with a reflection
symmetry was proposed. To construct a $3d$ SPT state with
reflection $\mathcal{P}$, one can start with a $2d$ system on the
reflection-invariant plane. The $3d$ SPT state can always be
constructed by stacking layers of $2d$ SPT states on the
reflection plane. But even when the $2d$ SPT state is a nontrivial
SPT state, it does not guarantee that the $3d$ SPT state is
nontrivial, more detailed analysis of the procedure of stacking is
demanded.

\begin{figure}
\includegraphics[width=0.8\linewidth]{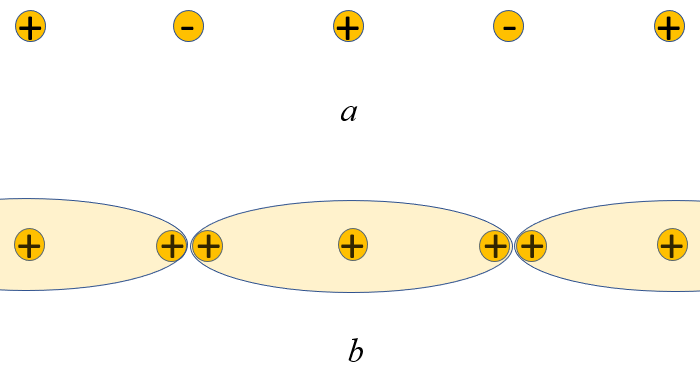}
\caption{The coupled wire construction for the symmetric gapped
edge state with $n=3$. Each wire is the boundary state of a $2d$
layer, and the wire with index $(-1)$ can be adiabatically
deformed into a wire with index $(+2)$ under interaction without
any transition in the $2d$ layer, due to the $\mathbb{Z}_3$
classification of the layer. Hence the wires can be regrouped and
gapped out by interactions that preserves all the symmetries.}
\label{wire2}
\end{figure}

\begin{figure}
\includegraphics[width=0.8\linewidth]{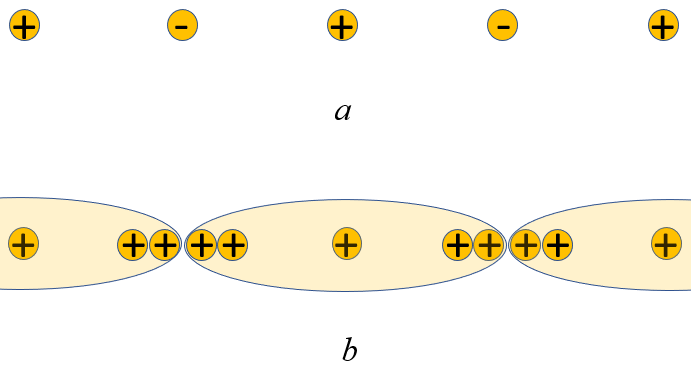}
\caption{Illustration of the symmetric gapped edge state with $n =
5$. Each wire with index $(-1)$ can be deformed into a wire with
index $(+4)$ under interaction, and again the wires can be
regrouped and gapped out by interactions that preserve all the
symmetries.} \label{wire3}
\end{figure}

The $2d$ boundary state of the $3d$ system can then be constructed
by stacking the $1d$ (wire) boundary state of the $2d$ layer SPT
states. This construction is often referred to as the coupled wire
construction. Two sides of the $1d$ wire are connected by
reflection $x \ra -x$, hence in the noninteracting limit this wire
can be viewed as the domain wall of the mass term $m(x)
\psi^\dagger \sigma^{20} \psi$ at the $2d$ boundary. The coupled
wire construction of the boundary states can be viewed as coupling
the domain wall states with oscillating sign of $m$ along the
$\hat{x}$ direction, and the domain walls are along the $\hat{y}$
direction. Let us assume that each domain has width $1$. At the
domain wall of $m(x)$, the domain wall wave functions are
eigenstates of $\sigma^{30} = (-1)^x$ for every other wire.

Since the wire states have eigenvalues $\sigma^{30} = \pm 1$, the
Hamiltonian and the symmetries project on the wire states as the
following: \beqn H(x)_{\mathrm{wire}} &=& \int dy \ (-1)^x
\psi^\dagger \ii \sigma^3
\partial_y \psi, \cr \cr \mathcal{P} &:& \psi(x) \rightarrow (-1)^x\psi(-x), \ \
\mathcal{T}: \psi \rightarrow (-1)^x \sigma^1 \psi, \cr\cr Z_n &:&
\psi \rightarrow \exp \left(\ii \frac{2\pi}{n} \sigma^{3} \right)
\psi. \eeqn By turning on tunnellings between the wires, one can
exactly reproduce the free fermion Hamiltonian of the $2d$
boundary state Eq.~\ref{2dedge}: \beqn H_{\mathrm{edge}} &=&
\sum_x \int dy \ (-1)^x \psi^\dagger \ii \sigma^3
\partial_y \psi + \ii t \psi^\dagger_x \sigma^0 \psi_{x + 1} +
H.c.
\cr\cr &\sim& \int d^2x \ \psi^\dagger ( \ii t \sigma^{10}
\partial_x + \ii \sigma^{33} \partial_y )\psi. \eeqn An extra
Pauli space emerges in the low energy theory because there are two
wires per unit cell in this wire construction. Here we turned on a
uniform tunnelling between neighboring wires. If instead a
staggered tunnelling $ \sum_x (-1)^x \ii t' \psi^\dagger_x
\sigma^0 \psi_{x + 1} + H.c. $ is turned on, the reflection
symmetry will be broken and a mass term $\psi^\dagger \sigma^{20}
\psi$ will be generated at low energy.

However, by turning on interaction, one can show that the entire
system is trivialized, even for a single copy of the $3d$ system.
In the following we will demonstrate this with $n = 3$, but this
argument can be generalized to any odd integer $n$. Let us give
index $(+1)$ to a $1d$ wire described by one pair of
counter-propagating $1d$ fermion modes, with fermion carrying
charge $\pm 1$ under symmetry $Z_3$ moving along the $\pm \hat{y}$
direction along the wire. First of all, quoting the results from
the previous section, each layer of the $2d$ SPT state has a
$\mathbb{Z}_3$ classification, hence each wire with index $(\pm
1)$ can be deformed continuously through interaction into two
pairs of counter-propagating $1d$ fermion modes, and the fermions
carrying $\pm 1$ $Z_3$ charges move along the $\mp \hat{y}$
direction. Or in other words a wire with index $(\pm 1)$ can be
continuously deformed in to a wire carrying index $(\mp 2)$
through interaction.

Now again, using the fact that the $2d$ interacting TI with
$(\U(1)\rtimes Z_2^T ) \times Z_3$ has a $\mathbb{Z}_3$
classification, we can group the wires as Fig.~\ref{wire2}$b$, and
turn on inter-wire interaction to gap out states along the wires,
and we can see that this arrangement preserves all the symmetries
including $\mathcal{P}$ and $\mathcal{T}$, and even translation
symmetry. This means that under interaction which preserves all
the symmetries, the boundary state of the $3d$ TI can be gapped
out by interaction, hence even a single copy of the $3d$ TI is
trivialized by interaction. The same construction can be
generalized to other odd integer $n$, for example the case with $n
= 5$ is illustrated in Fig.~\ref{wire3}$b$.

Now because the TI is trivialized by interaction, the bulk direct
trivial-to-topological transition in the noninteracting limit,
which is described by two massless $(3+1)d$ Dirac fermions,
becomes ``unnecessary" under strong enough interaction.

\subsection{Interacting TI: mapping to bosonic system}

Using the chiral boson language, we can still show that each wire
is equivalent to the edge state of a $2d$ bosonic symmetry
protected topological (bSPT) state, which can be shown by coupling
each wire to the boundary of a $2d$ bSPT state. We can still
describe this entire coupled $1d$ system using chiral bosons, and
this $1d$ system can be symmetrically gapped by the same
Lagrangian as $\mathcal{L}^{(2)}_{\mathrm{edge}}$
Eq.~\ref{Eq:QSH-BSPT_GappedInterface}. The $\mathcal{P}$,
$\mathcal{T}$ symmetries act on all the chiral boson fields as
\beqn \mathcal{P} &:& \phi_{1,2} \rightarrow \phi_{1,2} + \frac{1
+ (-1)^x}{2} \pi, \cr \cr && \varphi \rightarrow \varphi, \ \ \
\theta \rightarrow \theta; \cr\cr \mathcal{T} &:& \phi_{1,2}
\rightarrow - \phi_{2,1} + \frac{1 + (-1)^x}{2} \pi, \cr \cr &&
\varphi \rightarrow - \varphi, \ \ \ \theta \rightarrow \theta.
\eeqn $\exp(\ii\varphi)$ and $\exp(\ii\theta)$ are still local
bosons for odd integer $n$, $i.e.$ they are bound state of local
fermions.

There are standard formalisms of describing and constructing the
potentially nontrivial $3d$ bSPT state, such as the group
cohomology~\cite{wenspt,wenspt2}, and the effective nonlinear
sigma model with a topological
$\Theta-$term~\cite{senthilashvin,xusenthil,xuclass}. The bulk
bosonic state can be described by the action \beqn
\mathcal{S}_{\mathrm{bulk}} &=& \int d^3x d\tau \ \frac{1}{g}
(\partial_\mu \vect{n})^2 \cr\cr &+& \frac{ \Theta \ii}{\Omega_4}
\epsilon_{abcde} n^a \partial_x n^b
\partial_y n^c \partial_z n^d \partial_\tau n^e, \label{nlsm}\eeqn
where $\Theta = 2\pi$, $\Omega_4$ is the volume of a
four-dimensional sphere with unit radius. The unit-length
five-component vector field $\vect{n}(\vect{r}, \tau)$ can be
parameterized as \beqn (n_1, n_2) &=&
\cos(\beta)\cos(\gamma)\left(\cos(\varphi), \
\sin(\varphi)\right), \cr\cr (n_3, n_4) &=&
\cos(\beta)\sin(\gamma)\left(\cos(\theta), \ \sin(\theta)\right),
\cr\cr n_5 &=& \sin(\beta); \eeqn Under symmetries, the vector
$\vect{n}(\vect{r}, \tau)$ transform as
 \beqn \U(1) &:& \varphi \rightarrow \varphi + 2\alpha, \ \ \theta
 \rightarrow \theta, \ \ \beta \ra \beta \cr\cr
 Z_n &:& \varphi \rightarrow \varphi, \ \ \theta \rightarrow \theta - 2\pi
 /n, \ \ \beta \ra \beta
 \cr\cr \mathcal{T} &:& \varphi \ra - \varphi, \ \ \theta \ra
 \theta, \ \ \beta \ra \beta
 \cr\cr \mathcal{P} &:& \varphi \ra \varphi, \ \ \theta \ra
 \theta, \ \ \beta \ra - \beta.
 \eeqn
Hence $n_5$ is invariant under $(\U(1) \rtimes Z_2^T) \times Z_n$
symmetry, but odd under reflection $\mathcal{P}$. The action
Eq.~\ref{nlsm} is invariant under all the symmetries.

Based on the CPT theorem, we can replace a field theory with
$\mathcal{P}$ and $\mathcal{T}$ symmetry, by the $\mathcal{C}$ and
$\mathcal{T}$ symmetry, and $\mathcal{C}$ acts on the fields as
\beqn \mathcal{C}: \varphi \ra - \varphi, \ \ \ n_5 \rightarrow -
n_5. \eeqn Thus we can view $b \sim \cos(\varphi) + \ii
\sin(\varphi)$ as a bosonic rotor field, and $n_5$ is the density
of the boson. And $\mathcal{C}$ is the particle-hole
transformation of the bosonic rotor field. Now all the symmetries
are internal symmetries of the field theory Eq.~\ref{nlsm}.

Now as a consistency check we need to show that the bSPT is
actually trivial. This can be shown using the same method as
Ref.~\onlinecite{xulsm}. We can first embed the $Z_n$ symmetry
into another $\U(1)_s$ symmetry, and $\cos(\theta) + \ii
\sin(\theta)$ becomes a rotor under the $\U(1)_s$ symmetry. Then
the $3d$ bSPT can be understood using the ``decorated vortex"
picture. If we consider a vortex of line of $\theta$ (the vortex
configuration preserves the $\U(1)$, $Z_2^T$, and $\mathcal{C}$),
the action Eq.~\ref{nlsm} is reduced to the following $(1+1)d$
NLSM defined with a three component unit vector
$\tilde{\vect{n}}(x,\tau)$: \beqn \mathcal{S}_{\mathrm{1d}} = \int
dx d\tau \ \frac{1}{g}(\partial_\mu \tilde{\vect{n}})^2 +
\frac{\Theta \ii }{4\pi} \epsilon_{abc} \tilde{n}^a \partial_x
\tilde{n}^b
\partial_\tau \tilde{n}^c, \eeqn again $\Theta = 2\pi$.
The three component vector $\tilde{\vect{n}}$ is \beqn
\tilde{\vect{n}} = \left( \cos(\beta) \cos(\varphi), \ \cos(\beta)
\sin(\varphi), \ \sin(\beta) \right). \eeqn This implies that the
vortex of $\U(1)_s$ symmetry is decorated with a $1d$ Haldane
phase of $\tilde{\vect{n}}$~\cite{haldane1,haldane2,haldaneng},
which is known to have a $\mathbb{Z}_2$ classification. Thus the
$3d$ bSPT state can be constructed by first spontaneously breaking
the $\U(1)_s$ symmetry in the $3d$ bulk (developing a superfluid
phase), then decorate the vortex loop of this superfluid phase
with the Haldane phase described above, and then
proliferate/condense the decorated vortex loop.

Once we break $\U(1)_s$ down to $Z_n$ with odd integer $n$, this
decorated vortex picture will yield a trivial $3d$ bulk state.
This can be perceived by the fact that the $Z_n$ vortex loop has a
``$\mathbb{Z}_n$ classification", $i.e.$ the $n-$copies of $Z_n$
vortex loop is a trivial configuration in the space-time. The
$\mathbb{Z}_n$ classification with odd $n$ is incompatible with
the $\mathbb{Z}_2$ classification of the Haldane phase decorated
on the vortex loop.

More explicitly, one can demonstrate that the $2d$ boundary of
Eq.~\ref{nlsm} can be symmetrically gapped out without any
degeneracy with odd integer $n$. Again, let us start with the
$\U(1)_s$ symmetry, a single vortex of $\U(1)_s$ on the $2d$
boundary is the termination of the vortex line in the bulk, which
carries the $0d$ boundary state of the Haldane phase discussed
above, and due to the $\mathbb{Z}_2$ classification of the Haldane
phase, a double vortex of $\U(1)_s$ will carry trivial quantum
number, and hence can condense without breaking any symmetry.

After condensing the double vortex, the $2d$ boundary becomes a
$Z_2$ topological order, whose bosonic $e$ and $m$ anyon
excitations carry fractional quantum numbers. The $e$ excitation
is the remnant of the single vortex of $\U(1)_s$ symmetry after
condensing the double vortices, which carries a projective
representation of $\U(1) \rtimes Z_2^T$ and $\U(1) \rtimes
\mathcal{C}$. Here we pay particular attention to the $m$
excitation, which should carry half-charge of $\U(1)_s$, or
half-charge of $Z_n$ if we break $\U(1)_s$ down to its subgroup
$Z_n$. Hence under the $Z_n$ transformation, the $m$ excitation
acquires a phase factor \beqn Z_n : \Psi_m \ra \exp\left(\ii
\frac{2\pi}{2n}\right) \Psi_m. \eeqn Now consider a $n-$body bound
state of $\Psi_m$, let us denote it as $B \sim (\Psi_m)^n$. Under
the $Z_n$ symmetry, $B$ transforms as $B \ra - B $. And because
$n$ is an odd integer, $B \sim (\Psi_m)^n$ still carries $Z_2$
gauge charge$-1$, and the $Z_n$ transformation can be cancelled by
a $Z_2$ gauge transformation. $B$ can also be viewed as the bound
state between a single $\Psi_m$ and $(n-1)/2$ copies of the local
boson $e^{\ii \theta}$. Condensing $B$ at the $2d$ boundary does
not break any symmetry, and it confines the nontrivial anyons,
hence the boundary is driven into a fully gapped symmetric state
without degeneracy. This completes the argument that the bosonic
SPT state is actually trivial, which is consistent with our
conclusions in the previous subsections.

\section{Generic Unnecessary QCP in a $3d$ TSC}

An insulator has electron number conservation, hence it must have
a $\U(1)$ symmetry; while a superconductor breaks the particle
number conservation. In this section we discuss a superconductor
with $(Z_{2n} \rtimes Z_4^T)/Z_2$ symmetry, with an odd integer $n
> 1$. First of all, let us clarify the notation. The $Z_4^T$
stands for a time-reversal symmetry with $\mathcal{T}^2 = -1$,
$\mathcal{T}^4 = +1$. But we need to mod out the common $Z_2$
subgroup of both $Z_{2n}$ and $Z_4^T$. At the free fermion level,
there is a nontrivial TSC with such symmetry, whose $2d$ edge
state Hamiltonian and the symmetry transformation is \beqn
H_{\mathrm{edge}} &=& \int d^2x \ \psi^\dagger( \ii \sigma^1
\partial_x + \ii \sigma^3
\partial_y )\psi, \cr\cr Z_{2n} &:& \psi \ra \exp\left( \ii \frac{2\pi}{2n} \right)
\psi, \cr\cr \mathcal{T} &:& \psi \ra \ii \sigma^2 \psi.
\label{edge2}\eeqn Apparently with these symmetries no fermion
bilinear term can be turned on at the boundary Hamiltonian which
gaps out the boundary spectrum for any integer $n > 1$. Hence at
the free fermion level, Eq.~\ref{edge2} describes the boundary
state of a nontrivial TSC.

Our goal is to study the fate of this TSC under interaction. The
techniques we used in the previous section, $i.e.$ the coupled
wire construction, is no longer obviously applicable to this case.
But apparently this TSC can be embedded into the TI in the AII
class with $(\U(1) \rtimes Z_4^T)/Z_2$ symmetry, we can study the
interaction effect by starting with the $2d$ boundary topological
order of the AII TI constructed in
Ref.~\onlinecite{TI_fidkowski2,TI_qi,TI_senthil,TI_max}. This
boundary topological order is anomalous with the $\U(1)$ symmetry
of the AII class of TI, but we will show that this topological
order becomes nonanomalous and hence can be driven into a fully
symmetric gapped nondegenerate state, once $\U(1)$ is broken down
to $Z_{2n}$.

According to (for example) Ref.~\onlinecite{TI_max}, the boundary
of the AII TI can be driven into a topological order with in total
$48$ anyons (not including the electron itself). This topological
order can be constructed by first driving the boundary into a
superconductor by condensing Cooper pair $\psi^t \sigma^2 \psi$,
which spontaneously breaks the $\U(1)$ symmetry down to its $Z_2$
subgroup (the fermion parity of the electrons). Then the
symmetries an be restored by condensing the $8\pi$ vortex of the
superconductor, $i.e.$ eight fold bound state of the elementary
vortex of the superconductor, and the $8\pi$ vortex is a boson.
Within these anyons there is a charge $1/4$ boson $b$, which can
be viewed as a $1/8$ ``parton" of the Cooper pair. Now if we break
the $\U(1)$ down to $Z_{2n}$ symmetry with odd integer $n$, under
the $Z_{2n}$ transformation, this boson transforms as \beqn
Z_{2n}: b \rightarrow \exp\left(\ii \frac{2\pi}{8n} \right) b.
\eeqn $b$ is also coupled to a gauge field, and under the gauge
transformation, \beqn \mathrm{Gauge}: b \ra \exp \left(\ii \frac{k
\pi}{4}\right) b \eeqn with any integer $k$.

Now let us form a bound state of $b$ and $(n^2 - 1)/8$ copies of
Cooper pair $\psi^t \ii \sigma^2 \psi$: \beqn B \sim b \times
(\psi^t \ii \sigma^2 \psi)^{(n^2 - 1)/8}, \eeqn notice that for
odd integer $n
> 1$, $n^2 - 1$ is always an {\it integer multiple} of $8$. Then under
the $Z_{2n}$ symmetry, $B$ transforms as \beqn Z_{2n}: B &\ra&
\exp \left(\ii \frac{2\pi}{8n} + \ii 2 \times \frac{2\pi}{2n}
\times \frac{n^2 - 1}{8} \right) B \cr\cr &=& \left(\ii \frac{n
\pi}{4}\right) B. \eeqn This implies that the symmetry
transformation on $B$ can be cancelled by a gauge transformation.
$b$ and hence $B$ are both invariant under time-reversal.

Also, Ref.~\onlinecite{TI_max} demonstrated that $B$ has
nontrivial statistics with many of the anyons including the
nonabelian Ising anyon ($B$ does not carry any gauge independent
global quantum numbers, and it carries the same gauge charge as
$b$, because Cooper pairs are gauge neutral). This implies that
condensing $B$ would preserve all the symmetries, and confine all
the nontrivial anyons, $i.e.$ the condensate of $B$ is a fully
gapped symmetric $2d$ boundary state without ground state
degeneracy. In the condensate of $B$, $b$ can be identified as
multiples of Cooper pair $(\psi^t \ii \sigma^2 \psi)^{(n^2 -
1)/8}$.

There is also one deconfined neutral fermion $f$ that braids
trivially with boson $B$ and $b$, but this fermion $f$ is not a
fractionalized anyon in the condensate of $B$. $f$ can be
identified as the bound state of the original local fermion $\psi$
and multiple of $b$~\cite{TI_max}. Then when we break $\U(1)$ down
to $Z_{2n}$ with odd integer $n$, in the condensate of $B$, $f$
can be viewed as the bound state of $\psi$ and multiple of Cooper
pairs, which is also a Kramers doublet local fermion. This implies
that, once we break $\U(1)$ to $Z_{2n}$ with odd integer $n$,
interaction trivializes the TSC. Then the bulk
trivial-to-topological transition, which at the free fermion level
is formally described by a single massless Dirac fermion, becomes
a generic unnecessary QCP.

\section{Discussion}

In this work we propose two simple examples of generic unnecessary
QCPs, which are respectively described by {\it two} and {\it one}
massless $(3+1)d$ Dirac fermions, while the previously known
examples involve at least eight Dirac fermions. This result is
based on our analysis of classification of interacting $3d$ TI and
TSC. In both examples we demonstrated that the systems are
topological nontrivial without interaction, but are both totally
trivialized by local interactions. Local interaction is
perturbatively irrelevant at the noninteracting $(3+1)d$ Dirac
fermion fixed point, but a continuous route exists in the phase
diagram with strong enough interaction that connects the trivial
and topological phase of the TI and TSC in the noninteracting
limit (Fig.~\ref{AQCP}).

The $2d$ boundary Hamiltonian Eq.~\ref{2dedge} of the $3d$ TI in
our first example, as well as the transformation of the fermions
under symmetries are identical to the low energy theory of
spinless fermion at half-filling on the honeycomb lattice with
dominant nearest neighbor hopping. The fermion modes with
eigenvalue $\sigma^{03} = \pm 1$ correspond to the Dirac fermion
cones expanded at the two valleys of the Brillouin zone of the
honeycomb lattice. The $Z_3$ symmetry can be viewed as the
translation of the honeycomb lattice, and the reflection
$\mathcal{P}$ exchanges the A and B sublattices of the honeycomb
lattice. Our result supports that there exists a fully gapped and
symmetric state for interacting spinless fermions on the honeycomb
lattice, $i.e.$ there is no
Lieb-Shultz-Matthis~\cite{LSM,oshikawa,hastings} like theorem for
spinless fermions on the honeycomb lattice at half-filling with
translation and reflection symmetry, while this is only possible
under strong enough interaction.

There is another potentially interesting extension of our first
example. The trivial-to-topological transition of many bSPT
systems in $3d$, can be described by a $(3+1)d$ QCP with a
dynamical $\SU(2)$ gauge field coupled with two flavors of Dirac
fermions~\cite{deconfinedual}. This theory has a maximal $\SO(5)$
global symmetry. Breaking the $\SO(5)$ down to our symmetries
would permit more local quartic fermion terms in the Lagrangian.
The original trivial-to-topological transition of the bSPT system
is definitely ``unnecessary" because we know that this bSPT is
trivial once we break the $\SO(5)$ down to the symmetries
considered here. There is a possibility that this transition is
also a generic QCP which corresponds to a strongly interacting
conformal field theory. If this is the case, then the phase
diagram Fig.~\ref{AQCP} is even richer: there are two generic
unnecessary QCPs in the same phase diagram, but they belong to
different universality classes.

Chao-Ming Jian is supported by the Gordon and Betty Moore
Foundations EPiQS Initiative through Grant GBMF4304. Cenke Xu is
supported by NSF Grant No. DMR-1920434, and the David and Lucile
Packard Foundation. This work was performed in part at Aspen
Center for Physics, which is supported by National Science
Foundation grant PHY-1607611.

\bibliography{TRAN}

\begin{thebibliography}{59}
\expandafter\ifx\csname natexlab\endcsname\relax\def\natexlab#1{#1}\fi
\expandafter\ifx\csname bibnamefont\endcsname\relax
  \def\bibnamefont#1{#1}\fi
\expandafter\ifx\csname bibfnamefont\endcsname\relax
  \def\bibfnamefont#1{#1}\fi
\expandafter\ifx\csname citenamefont\endcsname\relax
  \def\citenamefont#1{#1}\fi
\expandafter\ifx\csname url\endcsname\relax
  \def\url#1{\texttt{#1}}\fi
\expandafter\ifx\csname urlprefix\endcsname\relax\def\urlprefix{URL }\fi
\providecommand{\bibinfo}[2]{#2}
\providecommand{\eprint}[2][]{\url{#2}}

\bibitem[{\citenamefont{WEN}(1990)}]{wenreviewtopo}
\bibinfo{author}{\bibfnamefont{X.~G.} \bibnamefont{WEN}},
  \bibinfo{journal}{International Journal of Modern Physics B}
  \textbf{\bibinfo{volume}{04}}, \bibinfo{pages}{239} (\bibinfo{year}{1990}),
  \eprint{https://doi.org/10.1142/S0217979290000139},
  \urlprefix\url{https://doi.org/10.1142/S0217979290000139}.

\bibitem[{\citenamefont{Chen et~al.}(2013)\citenamefont{Chen, Gu, Liu, and
  Wen}}]{wenspt}
\bibinfo{author}{\bibfnamefont{X.}~\bibnamefont{Chen}},
  \bibinfo{author}{\bibfnamefont{Z.-C.} \bibnamefont{Gu}},
  \bibinfo{author}{\bibfnamefont{Z.-X.} \bibnamefont{Liu}}, \bibnamefont{and}
  \bibinfo{author}{\bibfnamefont{X.-G.} \bibnamefont{Wen}},
  \bibinfo{journal}{Phys. Rev. B} \textbf{\bibinfo{volume}{87}},
  \bibinfo{pages}{155114} (\bibinfo{year}{2013}).

\bibitem[{\citenamefont{Chen et~al.}(2012)\citenamefont{Chen, Gu, Liu, and
  Wen}}]{wenspt2}
\bibinfo{author}{\bibfnamefont{X.}~\bibnamefont{Chen}},
  \bibinfo{author}{\bibfnamefont{Z.-C.} \bibnamefont{Gu}},
  \bibinfo{author}{\bibfnamefont{Z.-X.} \bibnamefont{Liu}}, \bibnamefont{and}
  \bibinfo{author}{\bibfnamefont{X.-G.} \bibnamefont{Wen}},
  \bibinfo{journal}{Science} \textbf{\bibinfo{volume}{338}},
  \bibinfo{pages}{1604} (\bibinfo{year}{2012}).

\bibitem[{\citenamefont{Senthil
  et~al.}(2004{\natexlab{a}})\citenamefont{Senthil, Vishwanath, Balents,
  Sachdev, and Fisher}}]{deconfine1}
\bibinfo{author}{\bibfnamefont{T.}~\bibnamefont{Senthil}},
  \bibinfo{author}{\bibfnamefont{A.}~\bibnamefont{Vishwanath}},
  \bibinfo{author}{\bibfnamefont{L.}~\bibnamefont{Balents}},
  \bibinfo{author}{\bibfnamefont{S.}~\bibnamefont{Sachdev}}, \bibnamefont{and}
  \bibinfo{author}{\bibfnamefont{M.~P.~A.} \bibnamefont{Fisher}},
  \bibinfo{journal}{Science} \textbf{\bibinfo{volume}{303}},
  \bibinfo{pages}{1490} (\bibinfo{year}{2004}{\natexlab{a}}).

\bibitem[{\citenamefont{Senthil
  et~al.}(2004{\natexlab{b}})\citenamefont{Senthil, Balents, Sachdev,
  Vishwanath, and Fisher}}]{deconfine2}
\bibinfo{author}{\bibfnamefont{T.}~\bibnamefont{Senthil}},
  \bibinfo{author}{\bibfnamefont{L.}~\bibnamefont{Balents}},
  \bibinfo{author}{\bibfnamefont{S.}~\bibnamefont{Sachdev}},
  \bibinfo{author}{\bibfnamefont{A.}~\bibnamefont{Vishwanath}},
  \bibnamefont{and} \bibinfo{author}{\bibfnamefont{M.~P.~A.}
  \bibnamefont{Fisher}}, \bibinfo{journal}{Phys. Rev. B}
  \textbf{\bibinfo{volume}{70}}, \bibinfo{pages}{144407}
  (\bibinfo{year}{2004}{\natexlab{b}}).

\bibitem[{\citenamefont{Sandvik}(2007)}]{deconfinesandvik1}
\bibinfo{author}{\bibfnamefont{A.~W.} \bibnamefont{Sandvik}},
  \bibinfo{journal}{Phys. Rev. Lett.} \textbf{\bibinfo{volume}{98}},
  \bibinfo{pages}{227202} (\bibinfo{year}{2007}),
  \urlprefix\url{https://link.aps.org/doi/10.1103/PhysRevLett.98.227202}.

\bibitem[{\citenamefont{Shao et~al.}(2016)\citenamefont{Shao, Guo, and
  Sandvik}}]{deconfinesandvik2}
\bibinfo{author}{\bibfnamefont{H.}~\bibnamefont{Shao}},
  \bibinfo{author}{\bibfnamefont{W.}~\bibnamefont{Guo}}, \bibnamefont{and}
  \bibinfo{author}{\bibfnamefont{A.~W.} \bibnamefont{Sandvik}},
  \bibinfo{journal}{Science} \textbf{\bibinfo{volume}{352}},
  \bibinfo{pages}{213} (\bibinfo{year}{2016}), ISSN \bibinfo{issn}{0036-8075},
  \eprint{https://science.sciencemag.org/content/352/6282/213.full.pdf},
  \urlprefix\url{https://science.sciencemag.org/content/352/6282/213}.

\bibitem[{\citenamefont{Melko and Kaul}(2008)}]{deconfinemelko}
\bibinfo{author}{\bibfnamefont{R.~G.} \bibnamefont{Melko}} \bibnamefont{and}
  \bibinfo{author}{\bibfnamefont{R.~K.} \bibnamefont{Kaul}},
  \bibinfo{journal}{Phys. Rev. Lett.} \textbf{\bibinfo{volume}{100}},
  \bibinfo{pages}{017203} (\bibinfo{year}{2008}),
  \urlprefix\url{https://link.aps.org/doi/10.1103/PhysRevLett.100.017203}.

\bibitem[{\citenamefont{Wang et~al.}(2017)\citenamefont{Wang, Nahum, Metlitski,
  Xu, and Senthil}}]{deconfinedual}
\bibinfo{author}{\bibfnamefont{C.}~\bibnamefont{Wang}},
  \bibinfo{author}{\bibfnamefont{A.}~\bibnamefont{Nahum}},
  \bibinfo{author}{\bibfnamefont{M.~A.} \bibnamefont{Metlitski}},
  \bibinfo{author}{\bibfnamefont{C.}~\bibnamefont{Xu}}, \bibnamefont{and}
  \bibinfo{author}{\bibfnamefont{T.}~\bibnamefont{Senthil}},
  \bibinfo{journal}{Phys. Rev. X} \textbf{\bibinfo{volume}{7}},
  \bibinfo{pages}{031051} (\bibinfo{year}{2017}),
  \urlprefix\url{https://link.aps.org/doi/10.1103/PhysRevX.7.031051}.

\bibitem[{\citenamefont{Qin et~al.}(2017)\citenamefont{Qin, He, You, Lu, Sen,
  Sandvik, Xu, and Meng}}]{deconfinedualnumeric}
\bibinfo{author}{\bibfnamefont{Y.~Q.} \bibnamefont{Qin}},
  \bibinfo{author}{\bibfnamefont{Y.-Y.} \bibnamefont{He}},
  \bibinfo{author}{\bibfnamefont{Y.-Z.} \bibnamefont{You}},
  \bibinfo{author}{\bibfnamefont{Z.-Y.} \bibnamefont{Lu}},
  \bibinfo{author}{\bibfnamefont{A.}~\bibnamefont{Sen}},
  \bibinfo{author}{\bibfnamefont{A.~W.} \bibnamefont{Sandvik}},
  \bibinfo{author}{\bibfnamefont{C.}~\bibnamefont{Xu}}, \bibnamefont{and}
  \bibinfo{author}{\bibfnamefont{Z.~Y.} \bibnamefont{Meng}},
  \bibinfo{journal}{Phys. Rev. X} \textbf{\bibinfo{volume}{7}},
  \bibinfo{pages}{031052} (\bibinfo{year}{2017}),
  \urlprefix\url{https://link.aps.org/doi/10.1103/PhysRevX.7.031052}.

\bibitem[{\citenamefont{Bi and Senthil}(2019)}]{bisenthil}
\bibinfo{author}{\bibfnamefont{Z.}~\bibnamefont{Bi}} \bibnamefont{and}
  \bibinfo{author}{\bibfnamefont{T.}~\bibnamefont{Senthil}},
  \bibinfo{journal}{Phys. Rev. X} \textbf{\bibinfo{volume}{9}},
  \bibinfo{pages}{021034} (\bibinfo{year}{2019}),
  \urlprefix\url{https://link.aps.org/doi/10.1103/PhysRevX.9.021034}.

\bibitem[{\citenamefont{Fidkowski and Kitaev}(2010)}]{fidkowski1}
\bibinfo{author}{\bibfnamefont{L.}~\bibnamefont{Fidkowski}} \bibnamefont{and}
  \bibinfo{author}{\bibfnamefont{A.}~\bibnamefont{Kitaev}},
  \bibinfo{journal}{Phys. Rev. B} \textbf{\bibinfo{volume}{81}},
  \bibinfo{pages}{134509} (\bibinfo{year}{2010}).

\bibitem[{\citenamefont{Fidkowski and Kitaev}(2011)}]{fidkowski2}
\bibinfo{author}{\bibfnamefont{L.}~\bibnamefont{Fidkowski}} \bibnamefont{and}
  \bibinfo{author}{\bibfnamefont{A.}~\bibnamefont{Kitaev}},
  \bibinfo{journal}{Phys. Rev. B} \textbf{\bibinfo{volume}{83}},
  \bibinfo{pages}{075103} (\bibinfo{year}{2011}).

\bibitem[{\citenamefont{Fidkowski et~al.}(2013)\citenamefont{Fidkowski, Chen,
  and Vishwanath}}]{chenhe3B}
\bibinfo{author}{\bibfnamefont{L.}~\bibnamefont{Fidkowski}},
  \bibinfo{author}{\bibfnamefont{X.}~\bibnamefont{Chen}}, \bibnamefont{and}
  \bibinfo{author}{\bibfnamefont{A.}~\bibnamefont{Vishwanath}},
  \bibinfo{journal}{Phys. Rev. X} \textbf{\bibinfo{volume}{3}},
  \bibinfo{pages}{041016} (\bibinfo{year}{2013}).

\bibitem[{\citenamefont{Wang and Senthil}(2014)}]{senthilhe3}
\bibinfo{author}{\bibfnamefont{C.}~\bibnamefont{Wang}} \bibnamefont{and}
  \bibinfo{author}{\bibfnamefont{T.}~\bibnamefont{Senthil}},
  \bibinfo{journal}{Phys. Rev. B} \textbf{\bibinfo{volume}{89}},
  \bibinfo{pages}{195124} (\bibinfo{year}{2014}).

\bibitem[{\citenamefont{You and Xu}(2014)}]{youinversion}
\bibinfo{author}{\bibfnamefont{Y.-Z.} \bibnamefont{You}} \bibnamefont{and}
  \bibinfo{author}{\bibfnamefont{C.}~\bibnamefont{Xu}}, \bibinfo{journal}{Phys.
  Rev. B} \textbf{\bibinfo{volume}{90}}, \bibinfo{pages}{245120}
  (\bibinfo{year}{2014}).

\bibitem[{\citenamefont{Qi}(2013)}]{qiz8}
\bibinfo{author}{\bibfnamefont{X.-L.} \bibnamefont{Qi}}, \bibinfo{journal}{New
  J. Phys.} \textbf{\bibinfo{volume}{15}}, \bibinfo{pages}{065002}
  (\bibinfo{year}{2013}).

\bibitem[{\citenamefont{Ryu and Zhang}(2012)}]{zhangz8}
\bibinfo{author}{\bibfnamefont{S.}~\bibnamefont{Ryu}} \bibnamefont{and}
  \bibinfo{author}{\bibfnamefont{S.-C.} \bibnamefont{Zhang}},
  \bibinfo{journal}{Phys. Rev. B} \textbf{\bibinfo{volume}{85}},
  \bibinfo{pages}{245132} (\bibinfo{year}{2012}).

\bibitem[{\citenamefont{Yao and Ryu}(2013)}]{yaoz8}
\bibinfo{author}{\bibfnamefont{H.}~\bibnamefont{Yao}} \bibnamefont{and}
  \bibinfo{author}{\bibfnamefont{S.}~\bibnamefont{Ryu}},
  \bibinfo{journal}{Phys. Rev. B} \textbf{\bibinfo{volume}{88}},
  \bibinfo{pages}{064507} (\bibinfo{year}{2013}).

\bibitem[{\citenamefont{Gu and Levin}(2013)}]{levinguz8}
\bibinfo{author}{\bibfnamefont{Z.-C.} \bibnamefont{Gu}} \bibnamefont{and}
  \bibinfo{author}{\bibfnamefont{M.}~\bibnamefont{Levin}},
  \bibinfo{journal}{arXiv:1304.4569}  (\bibinfo{year}{2013}).

\bibitem[{\citenamefont{Liu et~al.}(2018)\citenamefont{Liu, Vishwanath, and
  Khalaf}}]{ashvinzn}
\bibinfo{author}{\bibfnamefont{S.}~\bibnamefont{Liu}},
  \bibinfo{author}{\bibfnamefont{A.}~\bibnamefont{Vishwanath}},
  \bibnamefont{and} \bibinfo{author}{\bibfnamefont{E.}~\bibnamefont{Khalaf}},
  \bibinfo{journal}{arXiv:1809.01636}  (\bibinfo{year}{2018}).

\bibitem[{\citenamefont{Wu et~al.}(2019)\citenamefont{Wu, Xu, Jian, and
  Xu}}]{vci}
\bibinfo{author}{\bibfnamefont{X.-C.} \bibnamefont{Wu}},
  \bibinfo{author}{\bibfnamefont{Y.}~\bibnamefont{Xu}},
  \bibinfo{author}{\bibfnamefont{C.-M.} \bibnamefont{Jian}}, \bibnamefont{and}
  \bibinfo{author}{\bibfnamefont{C.}~\bibnamefont{Xu}},
  \bibinfo{journal}{arXiv:1906.07191}  (\bibinfo{year}{2019}).

\bibitem[{\citenamefont{Zhang et~al.}(2019{\natexlab{a}})\citenamefont{Zhang,
  Mao, Cao, Jarillo-Herrero, and Senthil}}]{band3}
\bibinfo{author}{\bibfnamefont{Y.-H.} \bibnamefont{Zhang}},
  \bibinfo{author}{\bibfnamefont{D.}~\bibnamefont{Mao}},
  \bibinfo{author}{\bibfnamefont{Y.}~\bibnamefont{Cao}},
  \bibinfo{author}{\bibfnamefont{P.}~\bibnamefont{Jarillo-Herrero}},
  \bibnamefont{and} \bibinfo{author}{\bibfnamefont{T.}~\bibnamefont{Senthil}},
  \bibinfo{journal}{Phys. Rev. B} \textbf{\bibinfo{volume}{99}},
  \bibinfo{pages}{075127} (\bibinfo{year}{2019}{\natexlab{a}}),
  \urlprefix\url{https://link.aps.org/doi/10.1103/PhysRevB.99.075127}.

\bibitem[{\citenamefont{Chittari et~al.}(2019)\citenamefont{Chittari, Chen,
  Zhang, Wang, and Jung}}]{band5}
\bibinfo{author}{\bibfnamefont{B.~L.} \bibnamefont{Chittari}},
  \bibinfo{author}{\bibfnamefont{G.}~\bibnamefont{Chen}},
  \bibinfo{author}{\bibfnamefont{Y.}~\bibnamefont{Zhang}},
  \bibinfo{author}{\bibfnamefont{F.}~\bibnamefont{Wang}}, \bibnamefont{and}
  \bibinfo{author}{\bibfnamefont{J.}~\bibnamefont{Jung}},
  \bibinfo{journal}{Phys. Rev. Lett.} \textbf{\bibinfo{volume}{122}},
  \bibinfo{pages}{016401} (\bibinfo{year}{2019}),
  \urlprefix\url{https://link.aps.org/doi/10.1103/PhysRevLett.122.016401}.

\bibitem[{\citenamefont{Lee et~al.}(2019)\citenamefont{Lee, Khalaf, Liu, Liu,
  Hao, Kim, and Vishwanath}}]{TDBGt}
\bibinfo{author}{\bibfnamefont{J.~Y.} \bibnamefont{Lee}},
  \bibinfo{author}{\bibfnamefont{E.}~\bibnamefont{Khalaf}},
  \bibinfo{author}{\bibfnamefont{S.}~\bibnamefont{Liu}},
  \bibinfo{author}{\bibfnamefont{X.}~\bibnamefont{Liu}},
  \bibinfo{author}{\bibfnamefont{Z.}~\bibnamefont{Hao}},
  \bibinfo{author}{\bibfnamefont{P.}~\bibnamefont{Kim}}, \bibnamefont{and}
  \bibinfo{author}{\bibfnamefont{A.}~\bibnamefont{Vishwanath}},
  \bibinfo{journal}{arXiv:1903.08130}  (\bibinfo{year}{2019}).

\bibitem[{\citenamefont{Liu et~al.}(2019{\natexlab{a}})\citenamefont{Liu, Ma,
  Gao, and Dai}}]{dai2}
\bibinfo{author}{\bibfnamefont{J.}~\bibnamefont{Liu}},
  \bibinfo{author}{\bibfnamefont{Z.}~\bibnamefont{Ma}},
  \bibinfo{author}{\bibfnamefont{J.}~\bibnamefont{Gao}}, \bibnamefont{and}
  \bibinfo{author}{\bibfnamefont{X.}~\bibnamefont{Dai}},
  \bibinfo{journal}{arXiv:1903.10419}  (\bibinfo{year}{2019}{\natexlab{a}}).

\bibitem[{\citenamefont{Chen et~al.}(2019)\citenamefont{Chen, Sharpe, Fox,
  Zhang, Wang, Jiang, Lyu, Li, Watanabe, Taniguchi et~al.}}]{hall}
\bibinfo{author}{\bibfnamefont{G.}~\bibnamefont{Chen}},
  \bibinfo{author}{\bibfnamefont{A.~L.} \bibnamefont{Sharpe}},
  \bibinfo{author}{\bibfnamefont{E.~J.} \bibnamefont{Fox}},
  \bibinfo{author}{\bibfnamefont{Y.-H.} \bibnamefont{Zhang}},
  \bibinfo{author}{\bibfnamefont{S.}~\bibnamefont{Wang}},
  \bibinfo{author}{\bibfnamefont{L.}~\bibnamefont{Jiang}},
  \bibinfo{author}{\bibfnamefont{B.}~\bibnamefont{Lyu}},
  \bibinfo{author}{\bibfnamefont{H.}~\bibnamefont{Li}},
  \bibinfo{author}{\bibfnamefont{K.}~\bibnamefont{Watanabe}},
  \bibinfo{author}{\bibfnamefont{T.}~\bibnamefont{Taniguchi}},
  \bibnamefont{et~al.}, \bibinfo{journal}{arXiv:1905.06535}
  (\bibinfo{year}{2019}).

\bibitem[{\citenamefont{Sharpe et~al.}(2019)\citenamefont{Sharpe, Fox, Barnard,
  Finney, Watanabe, Taniguchi, Kastner, and Goldhaber-Gordon}}]{FM}
\bibinfo{author}{\bibfnamefont{A.~L.} \bibnamefont{Sharpe}},
  \bibinfo{author}{\bibfnamefont{E.~J.} \bibnamefont{Fox}},
  \bibinfo{author}{\bibfnamefont{A.~W.} \bibnamefont{Barnard}},
  \bibinfo{author}{\bibfnamefont{J.}~\bibnamefont{Finney}},
  \bibinfo{author}{\bibfnamefont{K.}~\bibnamefont{Watanabe}},
  \bibinfo{author}{\bibfnamefont{T.}~\bibnamefont{Taniguchi}},
  \bibinfo{author}{\bibfnamefont{M.~A.} \bibnamefont{Kastner}},
  \bibnamefont{and}
  \bibinfo{author}{\bibfnamefont{D.}~\bibnamefont{Goldhaber-Gordon}},
  \bibinfo{journal}{arXiv:1901.03520}  (\bibinfo{year}{2019}).

\bibitem[{\citenamefont{Bultinck et~al.}(2019)\citenamefont{Bultinck,
  Chatterjee, and Zaletel}}]{zaletel}
\bibinfo{author}{\bibfnamefont{N.}~\bibnamefont{Bultinck}},
  \bibinfo{author}{\bibfnamefont{S.}~\bibnamefont{Chatterjee}},
  \bibnamefont{and} \bibinfo{author}{\bibfnamefont{M.~P.}
  \bibnamefont{Zaletel}}, \bibinfo{journal}{arXiv:1901.08110}
  (\bibinfo{year}{2019}).

\bibitem[{\citenamefont{Zhang et~al.}(2019{\natexlab{b}})\citenamefont{Zhang,
  Mao, and Senthil}}]{zhangmao}
\bibinfo{author}{\bibfnamefont{Y.-H.} \bibnamefont{Zhang}},
  \bibinfo{author}{\bibfnamefont{D.}~\bibnamefont{Mao}}, \bibnamefont{and}
  \bibinfo{author}{\bibfnamefont{T.}~\bibnamefont{Senthil}},
  \bibinfo{journal}{arXiv:1901.08209}  (\bibinfo{year}{2019}{\natexlab{b}}).

\bibitem[{\citenamefont{Kim}(2019)}]{kimtalk}
\bibinfo{author}{\bibfnamefont{P.}~\bibnamefont{Kim}},
  \emph{\bibinfo{title}{{Ferromagnetic superconductivity in twisted double
  bilayer graphene}}},
  \bibinfo{howpublished}{\url{http://online.kitp.ucsb.edu/online/bands_m19/kim%
/}} (\bibinfo{year}{2019}), \bibinfo{note}{{T}alks at KITP, Jan 15, 2019}.

\bibitem[{\citenamefont{Shen et~al.}(2019)\citenamefont{Shen, Li, Wang, Zhao,
  Tang, Liu, Tian, Chu, Watanabe, Taniguchi et~al.}}]{TDBG1}
\bibinfo{author}{\bibfnamefont{C.}~\bibnamefont{Shen}},
  \bibinfo{author}{\bibfnamefont{N.}~\bibnamefont{Li}},
  \bibinfo{author}{\bibfnamefont{S.}~\bibnamefont{Wang}},
  \bibinfo{author}{\bibfnamefont{Y.}~\bibnamefont{Zhao}},
  \bibinfo{author}{\bibfnamefont{J.}~\bibnamefont{Tang}},
  \bibinfo{author}{\bibfnamefont{J.}~\bibnamefont{Liu}},
  \bibinfo{author}{\bibfnamefont{J.}~\bibnamefont{Tian}},
  \bibinfo{author}{\bibfnamefont{Y.}~\bibnamefont{Chu}},
  \bibinfo{author}{\bibfnamefont{K.}~\bibnamefont{Watanabe}},
  \bibinfo{author}{\bibfnamefont{T.}~\bibnamefont{Taniguchi}},
  \bibnamefont{et~al.}, \bibinfo{journal}{arXiv:1903.06952}
  (\bibinfo{year}{2019}).

\bibitem[{\citenamefont{Liu et~al.}(2019{\natexlab{b}})\citenamefont{Liu, Hao,
  Khalaf, Lee, Watanabe, Taniguchi, Vishwanath, and Kim}}]{TDBG2}
\bibinfo{author}{\bibfnamefont{X.}~\bibnamefont{Liu}},
  \bibinfo{author}{\bibfnamefont{Z.}~\bibnamefont{Hao}},
  \bibinfo{author}{\bibfnamefont{E.}~\bibnamefont{Khalaf}},
  \bibinfo{author}{\bibfnamefont{J.~Y.} \bibnamefont{Lee}},
  \bibinfo{author}{\bibfnamefont{K.}~\bibnamefont{Watanabe}},
  \bibinfo{author}{\bibfnamefont{T.}~\bibnamefont{Taniguchi}},
  \bibinfo{author}{\bibfnamefont{A.}~\bibnamefont{Vishwanath}},
  \bibnamefont{and} \bibinfo{author}{\bibfnamefont{P.}~\bibnamefont{Kim}},
  \bibinfo{journal}{arXiv:1903.08130}  (\bibinfo{year}{2019}{\natexlab{b}}).

\bibitem[{\citenamefont{Cao et~al.}(2019)\citenamefont{Cao, Rodan-Legrain,
  Rubies-Bigorda, Park, Watanabe, Taniguchi, and Jarillo-Herrero}}]{TDBG3}
\bibinfo{author}{\bibfnamefont{Y.}~\bibnamefont{Cao}},
  \bibinfo{author}{\bibfnamefont{D.}~\bibnamefont{Rodan-Legrain}},
  \bibinfo{author}{\bibfnamefont{O.}~\bibnamefont{Rubies-Bigorda}},
  \bibinfo{author}{\bibfnamefont{J.~M.} \bibnamefont{Park}},
  \bibinfo{author}{\bibfnamefont{K.}~\bibnamefont{Watanabe}},
  \bibinfo{author}{\bibfnamefont{T.}~\bibnamefont{Taniguchi}},
  \bibnamefont{and}
  \bibinfo{author}{\bibfnamefont{P.}~\bibnamefont{Jarillo-Herrero}},
  \bibinfo{journal}{arXiv:1903.08596}  (\bibinfo{year}{2019}).

\bibitem[{\citenamefont{Wen and Zee}(1992)}]{wenzee}
\bibinfo{author}{\bibfnamefont{X.~G.} \bibnamefont{Wen}} \bibnamefont{and}
  \bibinfo{author}{\bibfnamefont{A.}~\bibnamefont{Zee}},
  \bibinfo{journal}{Phys. Rev. B} \textbf{\bibinfo{volume}{46}},
  \bibinfo{pages}{2290} (\bibinfo{year}{1992}),
  \urlprefix\url{https://link.aps.org/doi/10.1103/PhysRevB.46.2290}.

\bibitem[{\citenamefont{Wen}(1990)}]{wenedge}
\bibinfo{author}{\bibfnamefont{X.~G.} \bibnamefont{Wen}},
  \bibinfo{journal}{Phys. Rev. Lett.} \textbf{\bibinfo{volume}{64}},
  \bibinfo{pages}{2206} (\bibinfo{year}{1990}),
  \urlprefix\url{https://link.aps.org/doi/10.1103/PhysRevLett.64.2206}.

\bibitem[{\citenamefont{WEN}(1992)}]{wenreview}
\bibinfo{author}{\bibfnamefont{X.-G.} \bibnamefont{WEN}},
  \bibinfo{journal}{International Journal of Modern Physics B}
  \textbf{\bibinfo{volume}{06}}, \bibinfo{pages}{1711} (\bibinfo{year}{1992}),
  \eprint{https://doi.org/10.1142/S0217979292000840},
  \urlprefix\url{https://doi.org/10.1142/S0217979292000840}.

\bibitem[{\citenamefont{Haldane}(1995)}]{nullhaldane}
\bibinfo{author}{\bibfnamefont{F.~D.~M.} \bibnamefont{Haldane}},
  \bibinfo{journal}{Phys. Rev. Lett.} \textbf{\bibinfo{volume}{74}},
  \bibinfo{pages}{2090} (\bibinfo{year}{1995}),
  \urlprefix\url{https://link.aps.org/doi/10.1103/PhysRevLett.74.2090}.

\bibitem[{\citenamefont{Levin}(2013)}]{nulllevin}
\bibinfo{author}{\bibfnamefont{M.}~\bibnamefont{Levin}},
  \bibinfo{journal}{Phys. Rev. X} \textbf{\bibinfo{volume}{3}},
  \bibinfo{pages}{021009} (\bibinfo{year}{2013}),
  \urlprefix\url{https://link.aps.org/doi/10.1103/PhysRevX.3.021009}.

\bibitem[{\citenamefont{Bi et~al.}(2017)\citenamefont{Bi, Zhang, You, Young,
  Balents, Liu, and Xu}}]{xugraphene}
\bibinfo{author}{\bibfnamefont{Z.}~\bibnamefont{Bi}},
  \bibinfo{author}{\bibfnamefont{R.}~\bibnamefont{Zhang}},
  \bibinfo{author}{\bibfnamefont{Y.-Z.} \bibnamefont{You}},
  \bibinfo{author}{\bibfnamefont{A.}~\bibnamefont{Young}},
  \bibinfo{author}{\bibfnamefont{L.}~\bibnamefont{Balents}},
  \bibinfo{author}{\bibfnamefont{C.-X.} \bibnamefont{Liu}}, \bibnamefont{and}
  \bibinfo{author}{\bibfnamefont{C.}~\bibnamefont{Xu}}, \bibinfo{journal}{Phys.
  Rev. Lett.} \textbf{\bibinfo{volume}{118}}, \bibinfo{pages}{126801}
  (\bibinfo{year}{2017}),
  \urlprefix\url{https://link.aps.org/doi/10.1103/PhysRevLett.118.126801}.

\bibitem[{\citenamefont{You et~al.}(2016)\citenamefont{You, Bi, Mao, and
  Xu}}]{spn}
\bibinfo{author}{\bibfnamefont{Y.-Z.} \bibnamefont{You}},
  \bibinfo{author}{\bibfnamefont{Z.}~\bibnamefont{Bi}},
  \bibinfo{author}{\bibfnamefont{D.}~\bibnamefont{Mao}}, \bibnamefont{and}
  \bibinfo{author}{\bibfnamefont{C.}~\bibnamefont{Xu}}, \bibinfo{journal}{Phys.
  Rev. B} \textbf{\bibinfo{volume}{93}}, \bibinfo{pages}{125101}
  (\bibinfo{year}{2016}),
  \urlprefix\url{http://link.aps.org/doi/10.1103/PhysRevB.93.125101}.

\bibitem[{\citenamefont{{You} et~al.}(2015)\citenamefont{{You}, {Bi},
  {Rasmussen}, {Cheng}, and {Xu}}}]{xufb}
\bibinfo{author}{\bibfnamefont{Y.-Z.} \bibnamefont{{You}}},
  \bibinfo{author}{\bibfnamefont{Z.}~\bibnamefont{{Bi}}},
  \bibinfo{author}{\bibfnamefont{A.}~\bibnamefont{{Rasmussen}}},
  \bibinfo{author}{\bibfnamefont{M.}~\bibnamefont{{Cheng}}}, \bibnamefont{and}
  \bibinfo{author}{\bibfnamefont{C.}~\bibnamefont{{Xu}}}, \bibinfo{journal}{New
  Journal of Physics} \textbf{\bibinfo{volume}{17}}, \bibinfo{eid}{075010}
  (\bibinfo{year}{2015}), \eprint{1404.6256}.

\bibitem[{\citenamefont{Song et~al.}(2017{\natexlab{a}})\citenamefont{Song,
  Huang, Fu, and Hermele}}]{hermelecrystal}
\bibinfo{author}{\bibfnamefont{H.}~\bibnamefont{Song}},
  \bibinfo{author}{\bibfnamefont{S.-J.} \bibnamefont{Huang}},
  \bibinfo{author}{\bibfnamefont{L.}~\bibnamefont{Fu}}, \bibnamefont{and}
  \bibinfo{author}{\bibfnamefont{M.}~\bibnamefont{Hermele}},
  \bibinfo{journal}{Phys. Rev. X} \textbf{\bibinfo{volume}{7}},
  \bibinfo{pages}{011020} (\bibinfo{year}{2017}{\natexlab{a}}),
  \urlprefix\url{https://link.aps.org/doi/10.1103/PhysRevX.7.011020}.

\bibitem[{\citenamefont{Lu and Vishwanath}(2012)}]{luashvin}
\bibinfo{author}{\bibfnamefont{Y.-M.} \bibnamefont{Lu}} \bibnamefont{and}
  \bibinfo{author}{\bibfnamefont{A.}~\bibnamefont{Vishwanath}},
  \bibinfo{journal}{Phys. Rev. B} \textbf{\bibinfo{volume}{86}},
  \bibinfo{pages}{125119} (\bibinfo{year}{2012}).

\bibitem[{\citenamefont{Song et~al.}(2017{\natexlab{b}})\citenamefont{Song,
  Huang, Fu, and Hermele}}]{hermelep}
\bibinfo{author}{\bibfnamefont{H.}~\bibnamefont{Song}},
  \bibinfo{author}{\bibfnamefont{S.-J.} \bibnamefont{Huang}},
  \bibinfo{author}{\bibfnamefont{L.}~\bibnamefont{Fu}}, \bibnamefont{and}
  \bibinfo{author}{\bibfnamefont{M.}~\bibnamefont{Hermele}},
  \bibinfo{journal}{Phys. Rev. X} \textbf{\bibinfo{volume}{7}},
  \bibinfo{pages}{011020} (\bibinfo{year}{2017}{\natexlab{b}}),
  \urlprefix\url{https://link.aps.org/doi/10.1103/PhysRevX.7.011020}.

\bibitem[{\citenamefont{Vishwanath and Senthil}(2013)}]{senthilashvin}
\bibinfo{author}{\bibfnamefont{A.}~\bibnamefont{Vishwanath}} \bibnamefont{and}
  \bibinfo{author}{\bibfnamefont{T.}~\bibnamefont{Senthil}},
  \bibinfo{journal}{Phys. Rev. X} \textbf{\bibinfo{volume}{3}},
  \bibinfo{pages}{011016} (\bibinfo{year}{2013}),
  \urlprefix\url{https://link.aps.org/doi/10.1103/PhysRevX.3.011016}.

\bibitem[{\citenamefont{Xu and Senthil}(2013)}]{xusenthil}
\bibinfo{author}{\bibfnamefont{C.}~\bibnamefont{Xu}} \bibnamefont{and}
  \bibinfo{author}{\bibfnamefont{T.}~\bibnamefont{Senthil}},
  \bibinfo{journal}{Phys. Rev. B} \textbf{\bibinfo{volume}{87}},
  \bibinfo{pages}{174412} (\bibinfo{year}{2013}).

\bibitem[{\citenamefont{Bi et~al.}(2015)\citenamefont{Bi, Rasmussen, Slagle,
  and Xu}}]{xuclass}
\bibinfo{author}{\bibfnamefont{Z.}~\bibnamefont{Bi}},
  \bibinfo{author}{\bibfnamefont{A.}~\bibnamefont{Rasmussen}},
  \bibinfo{author}{\bibfnamefont{K.}~\bibnamefont{Slagle}}, \bibnamefont{and}
  \bibinfo{author}{\bibfnamefont{C.}~\bibnamefont{Xu}}, \bibinfo{journal}{Phys.
  Rev. B} \textbf{\bibinfo{volume}{91}}, \bibinfo{pages}{134404}
  (\bibinfo{year}{2015}),
  \urlprefix\url{https://link.aps.org/doi/10.1103/PhysRevB.91.134404}.

\bibitem[{\citenamefont{Jian et~al.}(2018)\citenamefont{Jian, Bi, and
  Xu}}]{xulsm}
\bibinfo{author}{\bibfnamefont{C.-M.} \bibnamefont{Jian}},
  \bibinfo{author}{\bibfnamefont{Z.}~\bibnamefont{Bi}}, \bibnamefont{and}
  \bibinfo{author}{\bibfnamefont{C.}~\bibnamefont{Xu}}, \bibinfo{journal}{Phys.
  Rev. B} \textbf{\bibinfo{volume}{97}}, \bibinfo{pages}{054412}
  (\bibinfo{year}{2018}),
  \urlprefix\url{https://link.aps.org/doi/10.1103/PhysRevB.97.054412}.

\bibitem[{\citenamefont{Haldane}(1983{\natexlab{a}})}]{haldane1}
\bibinfo{author}{\bibfnamefont{F.~D.~M.} \bibnamefont{Haldane}},
  \bibinfo{journal}{Phys. Lett. A} \textbf{\bibinfo{volume}{93}},
  \bibinfo{pages}{464} (\bibinfo{year}{1983}{\natexlab{a}}).

\bibitem[{\citenamefont{Haldane}(1983{\natexlab{b}})}]{haldane2}
\bibinfo{author}{\bibfnamefont{F.~D.~M.} \bibnamefont{Haldane}},
  \bibinfo{journal}{Phys. Rev. Lett.} \textbf{\bibinfo{volume}{50}},
  \bibinfo{pages}{1153} (\bibinfo{year}{1983}{\natexlab{b}}).

\bibitem[{\citenamefont{Ng}(1994)}]{haldaneng}
\bibinfo{author}{\bibfnamefont{T.-K.} \bibnamefont{Ng}},
  \bibinfo{journal}{Phys. Rev. B} \textbf{\bibinfo{volume}{50}},
  \bibinfo{pages}{555} (\bibinfo{year}{1994}),
  \urlprefix\url{https://link.aps.org/doi/10.1103/PhysRevB.50.555}.

\bibitem[{\citenamefont{Chen et~al.}(2014)\citenamefont{Chen, Fidkowski, and
  Vishwanath}}]{TI_fidkowski2}
\bibinfo{author}{\bibfnamefont{X.}~\bibnamefont{Chen}},
  \bibinfo{author}{\bibfnamefont{L.}~\bibnamefont{Fidkowski}},
  \bibnamefont{and}
  \bibinfo{author}{\bibfnamefont{A.}~\bibnamefont{Vishwanath}},
  \bibinfo{journal}{Phys. Rev. B} \textbf{\bibinfo{volume}{89}},
  \bibinfo{pages}{165132} (\bibinfo{year}{2014}),
  \urlprefix\url{https://link.aps.org/doi/10.1103/PhysRevB.89.165132}.

\bibitem[{\citenamefont{Bonderson et~al.}(2013)\citenamefont{Bonderson, Nayak,
  and Qi}}]{TI_qi}
\bibinfo{author}{\bibfnamefont{P.}~\bibnamefont{Bonderson}},
  \bibinfo{author}{\bibfnamefont{C.}~\bibnamefont{Nayak}}, \bibnamefont{and}
  \bibinfo{author}{\bibfnamefont{X.-L.} \bibnamefont{Qi}}, \bibinfo{journal}{J.
  Stat. Mech.} p. \bibinfo{pages}{P09016} (\bibinfo{year}{2013}).

\bibitem[{\citenamefont{Wang et~al.}(2013)\citenamefont{Wang, Potter, and
  Senthil}}]{TI_senthil}
\bibinfo{author}{\bibfnamefont{C.}~\bibnamefont{Wang}},
  \bibinfo{author}{\bibfnamefont{A.~C.} \bibnamefont{Potter}},
  \bibnamefont{and} \bibinfo{author}{\bibfnamefont{T.}~\bibnamefont{Senthil}},
  \bibinfo{journal}{Phys. Rev. B} \textbf{\bibinfo{volume}{88}},
  \bibinfo{pages}{115137} (\bibinfo{year}{2013}),
  \urlprefix\url{https://link.aps.org/doi/10.1103/PhysRevB.88.115137}.

\bibitem[{\citenamefont{Metlitski et~al.}(2015)\citenamefont{Metlitski, Kane,
  and Fisher}}]{TI_max}
\bibinfo{author}{\bibfnamefont{M.~A.} \bibnamefont{Metlitski}},
  \bibinfo{author}{\bibfnamefont{C.~L.} \bibnamefont{Kane}}, \bibnamefont{and}
  \bibinfo{author}{\bibfnamefont{M.~P.~A.} \bibnamefont{Fisher}},
  \bibinfo{journal}{Phys. Rev. B} \textbf{\bibinfo{volume}{92}},
  \bibinfo{pages}{125111} (\bibinfo{year}{2015}),
  \urlprefix\url{https://link.aps.org/doi/10.1103/PhysRevB.92.125111}.

\bibitem[{\citenamefont{Lieb et~al.}(1961)\citenamefont{Lieb, Schultz, and
  Mattis}}]{LSM}
\bibinfo{author}{\bibfnamefont{E.~H.} \bibnamefont{Lieb}},
  \bibinfo{author}{\bibfnamefont{T.~D.} \bibnamefont{Schultz}},
  \bibnamefont{and} \bibinfo{author}{\bibfnamefont{D.~C.}
  \bibnamefont{Mattis}}, \bibinfo{journal}{Ann. Phys.}
  \textbf{\bibinfo{volume}{16}}, \bibinfo{pages}{407} (\bibinfo{year}{1961}).

\bibitem[{\citenamefont{Oshikawa}(2000)}]{oshikawa}
\bibinfo{author}{\bibfnamefont{M.}~\bibnamefont{Oshikawa}},
  \bibinfo{journal}{Phys. Rev. Lett.} \textbf{\bibinfo{volume}{84}},
  \bibinfo{pages}{1535} (\bibinfo{year}{2000}),
  \urlprefix\url{https://link.aps.org/doi/10.1103/PhysRevLett.84.1535}.

\bibitem[{\citenamefont{Hastings}(2004)}]{hastings}
\bibinfo{author}{\bibfnamefont{M.~B.} \bibnamefont{Hastings}},
  \bibinfo{journal}{Phys. Rev. B} \textbf{\bibinfo{volume}{69}},
  \bibinfo{pages}{104431} (\bibinfo{year}{2004}),
  \urlprefix\url{http://link.aps.org/doi/10.1103/PhysRevB.69.104431}.

\end{thebibliography}

\end{document}